\documentclass[submission,copyright,creativecommons]{eptcs}
\providecommand{\event}{FOCLASA 2012}

\usepackage{amsthm}
\usepackage[english]{babel}
\usepackage{breakurl}
\usepackage{cite}
\usepackage{graphicx}
\usepackage{listings}
\usepackage{localdef}
\usepackage{mathtools}
\usepackage{subfigure}
\usepackage{url}

\newcommand{\Bach}{{\sf Bach}}
\newcommand{\GAMMA}{{\sf GAMMA}}
\newcommand{\KLAIM}{{\sf KLAIM}}
\newcommand{\LIME}{{\sc LIME}}
\newcommand{\LINDA}{{\sc LINDA}}
\newcommand{\LOGOP}{{\sc LogOp}}
\newcommand{\TUCSON}{{\sf TuCSoN}}
\newcommand{\RESPECT}{{\sf ReSpecT}}

\title{Blackboard Rules for Coordinating
Context-aware Applications in Mobile Ad Hoc Networks}
\def\titlerunning{Coordination Through Blackboard Rules}

\author{
Jean-Marie Jacquet \qquad\qquad Isabelle Linden \qquad\qquad Mihail-Octavian
Staicu
	\institute{	University of Namur\\
				Namur, Belgium
			}
	\email{\quad jmj@info.fundp.ac.be \quad\qquad isabelle.linden@fundp.ac.be
	\quad\qquad msc@info.fundp.ac.be}
}

\def\authorrunning{J.-M. Jacquet, I. Linden, M.-O. Staicu}

\begin{document}

\maketitle

\begin{abstract}

Thanks to improvements in wireless communication technologies and
increasing computing power in hand-held devices, mobile ad hoc networks are
becoming an ever-more present reality. Coordination languages are expected to
become important means in supporting this type of interaction. To this extent
we argue the interest of the \Bach\ coordination language as a middleware that
can handle and react to context changes as well as cope with unpredictable
physical interruptions that occur in opportunistic network connections. More
concretely, our proposal is based on blackboard rules that model declaratively
the actions to be taken once the blackboard content reaches a predefined state,
but also that manage the engagement and disengagement of hosts and transient
sharing of blackboards. The idea of reactiveness has already been introduced in
previous work, but as will be appreciated by the reader, this article
presents a new perspective, more focused on a declarative setting.
\end{abstract}

\section{Introduction}
\label{sectIntroduction}

On the front-line of current research areas, mobility related technologies are
being developed in an exponential manner as a natural response to our
information needs, regardless of space or time. Such information is mostly
related to the context in which the user finds himself and usually dependent on
his current needs. In the field of mobility, applications must cope with
constant context changes in order to provide relevant information on demand.
With current development strongly tackling the mobile connectivity domain, we
can understand why the seminal equation
\begin{center}
	$programming~=~coordination~+~computation$~\cite{GC92}\\
\end{center}
remains of great relevance. \emph{Coordination} is thus an equal partner in the
\emph{programming} process as \emph{computation}. This is even an essential
element in the fields of distributed systems and concurrency. To be as generic as
possible, coordination acts as a middleware layer between a consumer and a
producer. It must place a demand to an entity able to solve it and afterwards
ensures that an answer is returned. Departing from these two ideas we can see
how \emph{mobility} and \emph{coordination} are key concepts for which proper
mechanisms of interaction are required. As such, our objectives in this paper
are to  tackle two important needs: supporting context awareness and solving
mobility issues in ad hoc networks related to the unpredictability of their
connections and topology. We propose a solution in the form of \emph{rules}
spanning over one or more device data spaces (henceforth \emph{blackboards}):
\begin{equation}
	in(a,X),~nin(b,Y)~~\longrightarrow~~in(c,W),~nin(d,Z)
\label{eqn1}
\end{equation}
with the intuitive meaning that as long as some information \emph{X} is present
on a blackboard \emph{a} and some information \emph{Y} is not present on
blackboard \emph{b}, then we can assume that some information \emph{W} will
become available on blackboard \emph{c} and some information \emph{Z} will no
longer be available on blackboard \emph{d}.

By providing two operational readings of such declarative rules, we aim at
obtaining both of these needed functionalities. First, by means of a
\emph{forward reading} (inferred from applying the rule from left to right), we
model the reaction to context changes occurring on the blackboards of mobile
devices. Intuitively, departing from the previous rule, a blackboard context is
defined in terms of information being present and/or absent and can model
a user's specific need. For example, the rule
\begin{center}
$in(a,\langle location,~Brussels \rangle),~nin(a,\langle weather,~sunny \rangle
)~~\longrightarrow~~in(a,\langle list,~museums \rangle)$
\end{center}
states that if a tourist is in Brussels and the weather is not sunny, he would
like to receive a list of available museums.
Second, through a \emph{backward reading} (inferred from using the rule from
right to left), we represent the links between blackboards that would model the
connection state of two devices. Let us assume the same tourist who is visiting
Brussels, owning a mobile device, throughout the city info posts can make
available a broad range of information consisting of weather data, nearby
parking places, public transportation, etc. As he walks around the
city, his mobile device roams between different such info posts and eventually
connects to these when coming within connection range, which can depend on the
wireless technology used for the connection. By exploiting the events raised at
the networking level, a backward reading rule of the form
$in(b,X)~~\longrightarrow~~in(a,X)$ is generated on the tourist's blackboard in
order to define the connection state of his mobile device (containing the
blackboard $a$) and the info post (containing the blackboard $b$). This rule
acts as a logical link and allows for the transient sharing of the two
blackboards, without the need to physically transfer the information between
the blackboards. In conjunction with this rule, the previous forward reading
rule can be resolved since the info post will provide the needed information
about the weather.

The rest of this paper is organized as follows.
Section~\ref{sectBBCoordRules} presents the concept of multiple blackboard
coordination through rules.
Section~\ref{sectImplLim} shows an operational implementation of the rules on
the \Bach\ language.
Section~\ref{sectTechDet} presents some technical and
architecture details of our mobile implementation.
Section~\ref{sectMobCoord} presents related work in the field of mobile
coordination and how our work relates to it.
Section~\ref{sectRelatedWork} presents some related work in terms of reactivity
and how our proposal is positioned with respect to these.
Section~\ref{sectConcl} draws the conclusions and presents the expectations for
future work.

\section{Blackboard coordination through rules}
\label{sectBBCoordRules}

Reactivity in coordination languages relies on the basic idea of triggering an
event when a predefined condition is met. The purpose is to enrich the
capabilities of classical tuple spaces in order to transform them from
simple tuple containers into context-aware entities, responsive to
the interactions performed by the agents. Extensions have been explored, either
by defining more complex triggering conditions or by taking a more complex set
of actions.

Our approach, based on the principle of reactivity, is used to model the
behavior of the system due to context changes on the tuple space and to handle
the execution of mobile agents in scenarios that involve physical mobility of
the hosts. Following the ideas exposed in \LIME~\cite{MPR06}, when two or more
devices come into each-other's range, their tuple spaces are merged in terms of
information that is publicly available. A hard to grasp notion is the one of
\emph{``close enough''} for two devices to be in range and thus to be connected.
Likewise, the idea of merging blackboards, although intuitive, is actually not
a clearcut one. Let us address these two points in turn. On the one hand,
closeness in practical terms is to be defined by the presence of a physical
communication link, be it wired or wireless, that would enable a communication
channel between the devices. From an abstract perspective, modeling such a
notion is very difficult to express; however our solution benefits from the
events raised by the operating system of the device and captured by the
middleware. On the other hand, merging the blackboards should not be seen
straightforward in the sense of transferring tuples over the network, but by
providing \emph{logical links}, as it was defined in~\cite{BJ96}. These are in
essence pointers that indicate the merging state with other blackboards. Taking
a declarative approach, having a local blackboard merged with a remote one
means that information is in some way available locally, assumed from the
presence or absence on other blackboards.

For example, the rule $in(b_{1},X)~~\longrightarrow~~in(b_{2},X)$ states in a
forward reading that any tuple $X$ present on $b_1$ should also be considered
present on $b_2$. By applying a backward reading we obtain a similar semantic
which asserts that $X$ is to be considered present on $b_2$ for as long as $X$
is present on $b_1$. The common ground between the two readings is represented
by the two expressions that comprise the rule, with respect to the
$\longrightarrow$: the \emph{left-hand side} (henceforth LHS) represents the
rule's activation condition and the \emph{right-hand side} (henceforth RHS)
represents the effect produced by the rule's activation. To make such a rule
more suitable for coping with context-awareness and transient sharing, we
develop the LHS and the RHS to a more generic form. To this end, we extend the
declarative equation~\ref{eqn1} into the more general expression represented in
equation~\ref{eqn2}. The context is represented by a sequence of template tuples
that should be present on (by use of the $in$ primitive) or absent from (by use
of $nin$ primitive) the blackboards. In the same way, the RHS can define the
tuples that should be considered present or absent on several blackboards.
There is no restriction for the blackboards defined in the LHS and the RHS in
the sense that they don't necessarily need to be the same. The more general
formal representation of the declarative rules is presented below:
\begin{eqnarray}
	\lefteqn{in(b_1,t_1),\cdots,in(b_m,t_m),nin(b_{m+1},t_{m+1}),\cdots,nin(b_n,t_n)}
	\nonumber \\
		&& \hspace*{10mm} \elmrarrow \hspace*{2mm}
		in(b_{n+1},t_{n+1}),\cdots,in(b_p,t_p),nin(b_{p+1},t_{p+1}),\cdots,nin(b_q,t_q)
\label{eqn2}
\end{eqnarray}
The meaning of this rule is that the presence of the tuples
$t_1,\cdots,t_m$ on the blackboards $b_1,\cdots,b_m$ respectively and the
absence of the tuples $t_{m+1},\cdots,t_n$ on the blackboards
$b_{m+1},\cdots,b_n$ respectively, implies the presence of tuples
$t_{n+1},\cdots,t_p$ on the blackboards $b_{n+1},\cdots,b_p$ respectively and
the absence of tuples $t_{p+1},\cdots,t_q$ on the blackboards
$b_{p+1},\cdots,b_q$ respectively. In the scenario of opportunistic mobile
ad hoc networks, the operational translation of the LHS requires the
acquiring of a lock on distributed blackboards, which is not feasible due to the
unpredictability of the connections. As we will see in
section~\ref{sectImplLimSubReadings}, the solution amounts to restricting a
rule's activation condition to a single blackboard.

The implicit behaviour of a rule upon activation is that information is consumed
from the LHS in order to obtain the effect defined in the RHS. One important
remark to be made is that for the $nin$ primitive the destructive behaviour
actually results in producing the template tuple on the respective blackboard.
However, when using the rules to represent the transient sharing between
blackboards such destructive behaviour is not desired. In order to preserve the
characteristics of a logical link, the content of the blackboards defined in the
rule must not be altered. To this extent we introduce a semantic notation in the
form of square brackets surrounding the primitives $in$ or $nin$, obtaining the
guarded variants $[in(b,t)]$ and $[nin(b,t)]$. In order to fine tune the effect
of a rule's activation, both the LHS and the RHS can have sets of simple
primitives or guarded ones.

By providing two operational readings of equation~\ref{eqn2} it is possible to
obtain two different functionalities: \emph{(i)} context awareness, by means of a
\emph{forward reading} and \emph{(ii)} coordination in mobile ad hoc networks,
by means of a \emph{backward reading}. The detailed description of each reading
will be provided in the next sections.

Note that other work, such as \TUCSON~\cite{OZ99} or \LIME~\cite{MPR01},
incorporate \emph{IF pattern THEN actions} mechanisms. However, as we will show
in section~\ref{sectImplLim}, our declarative approach allows to have a more
refined view. For the moment, let us simply stress that in contrast to such an
operational mechanism, we provide two multi directional views out of a
declarative implication.

\section{Language design}
\label{sectImplLim}

Having presented the declarative definition of the blackboard rules in
section~\ref{sectIntroduction}, we dedicate this section to describe how they
can be incorporated in the \Bach\ coordination language. With this aim, the
language is first briefly presented in subsection~\ref{sectImplLimSubBach}. In
subsection~\ref{sectImplLimSubReadings} we expand the idea of \emph{rule
reading} in order to model different functionalities and explain the constraints
imposed by MANETs. We conclude with subsection~\ref{sectImplLimSubExamples} by
presenting examples for each of the two readings. A section on formal
operational semantics has also been developed, but due to page limitations could
not be introduced in this version of the paper. The interested reader can refer
to the full version of the paper which is available online
\footnote{\url{http://info.fundp.ac.be/~msc/articles/FOCLASA_2012_full.pdf}}.

\subsection{\Bach}
\label{sectImplLimSubBach}
Introduced in~\cite{JL07a}, the \Bach\ coordination language is built
upon the principle of a central \emph{blackboard} (the equivalent of
Linda's \emph{tuple space}) represented by a shared memory space through
which agents can communicate. The interaction with the blackboard is
ensured by the use of four primitives: $tell$ for outputting information on
the blackboard, $ask$ for querying the presence of information, $get$ for
retrieving information and $nask$ for querying the absence of information. In
order to express more complex actions, the four primitives can be linked by
composition operators, namely: $";"$ for sequential execution, $"||"$ for
parallel execution and $"+"$ for nondeterministic choice execution.

We improve the language definition by adding a set of four similar primitives
for handling the rules over a network:
\[
	\mathit{tellr}(b,r) \hspace{10mm} \mathit{askr}(b,r) \hspace{10mm}
	\mathit{getr}(b,r) \hspace{10mm} \mathit{naskr}(b,r)
\]
having the respective semantics of adding a rule $r$ to a blackboard $b$,
querying the presence, retrieving, querying the absence.

\subsection{Forward and backward reading}
\label{sectImplLimSubReadings}

As illustrated by other declarative languages (Prolog~\cite{SS94},
Haskell~\cite{T99}, etc.), the declarative reading of the blackboard rules must
be completed by an operational one. To that aim let us now present in detail
the two readings of the blackboard rules introduced in equation~\ref{eqn2}.
\\
\\
\textbf{Forward reading}~($LHS \longrightarrow_f RHS$). A first reading, named
subsequently \emph{forward reading}, is obtained by reading the
general rule expression from left to right and by acting accordingly:
provided that the condition in the LHS is met, then for each possibility in
which the condition can be expressed from the blackboard's context the
statements in the RHS are made true. Operationally, this means the following:

\begin{itemize}
  \item For each tuple $t_i~(1 \leq i \leq m)$ added on $b_i$ and for any
  tuple $t_j$ not present on $b_j~(m+1 \leq j \leq n)$, then for any
  set of tuples $(t_1,\cdots,t_m)$ from the set of blackboards
  $(b_1,\cdots,b_m)$, the corresponding tuples $t_k~(n+1 \leq k
  \leq p)$ should be created on $b_k~(n+1 \leq k
  \leq p)$ and the corresponding tuples $t_l$ should be removed from $b_l~(p+1
  \leq l \leq q)$.
  \item For each tuple $t_i~(1 \leq i \leq m)$ removed from $b_i$ and for
  any tuple $t_j$ not present on $b_j~(m+1 \leq j \leq n)$, then for
  any set of tuples $(t_1,\cdots,t_m)$ from the set of blackboards
  $(b_1,\cdots,b_m)$, the corresponding tuples $t_k~(n+1 \leq k
  \leq p)$ should be created on $b_k~(n+1 \leq k
  \leq p)$ and the corresponding tuples $t_l$ should be removed from $b_l~(p+1
  \leq l \leq q)$.
\end{itemize}

The activation condition of a rule represents a context defined on
multiple blackboards. However, a full implementation of that general rule would
involve a costly mechanism and heavy network loading in order to check whether
the condition becomes active or not. Furthermore, in mobile ad hoc networks the
communication links between devices are unstable and unpredictable. In this
hypothesis, the wise choice is to restrict the context definition to only one
blackboard. As such, the operational forward reading is represented as:\\
\indent
$in(b_1,t_1), \cdots, in(b_1,t_m),nin(b_1,t_{m+1}), \cdots,
nin(b_1,t_n)$\\
\hspace*{10mm}$\elmrarrow_f$ \hspace*{2mm} $in(b_{n+1},t_{n+1}), \cdots,
in(b_p,t_p),nin(b_{p+1},t_{p+1}), \cdots, nin(b_q,t_q)$
\\
\\
\textbf{Backward reading}~($LHS \longrightarrow_b RHS$). A second reading is
obtained in a backward fashion by reading the general rule from right to
left. The behavior is also different from the forward reading since as long as
the conditions in the LHS are met, the statements in the RHS are verified. As such,

\begin{itemize}
  \item The presence of $t_k$ on $b_k~(n+1 \leq k \leq p)$ can be deduced from
  the presence of a set of tuples $(t_1,\cdots,t_m)$ on the set of blackboards
  $(b_1,\cdots,b_m)$ and the absence of a set of tuples $(t_{m+1},\cdots,t_n)$
  on the set of blackboards $(b_{m+1},\cdots,b_n)$.
  \item The absence of $t_l$ from $b_l~(p+1 \leq k \leq q)$ can be deduced from
  the presence of a set of tuples $(t_1,\cdots,t_m)$ on the set of blackboards
  $(b_1,\cdots,b_m)$ and the absence of a set of tuples $(t_{m+1},\cdots,t_n)$
  on the set of blackboards $(b_{m+1},\cdots,b_n)$.
\end{itemize}

For the backward reading, restrictions must be imposed in the RHS as well: only
one $in$ or $nin$ primitive is permitted. The reason is that the presence or
absence of a tuple on a local blackboard can be inferred from the context of
another blackboard, thus providing pointers to virtual tuples that can be used
in the evaluation of $ask$, $get$, $nask$ primitives. To this extent, the
operational backward reading has two possible forms:\\
\indent $in(b_1,t_1),\cdots, in(b_1,t_m),nin(b_1,t_{m+1}), \cdots, nin(b_1,t_n)$
	\hspace*{2mm}$\elmrarrow_b$ \hspace*{2mm} $in(b_2,t_{n+1})$\\
\indent $in(b_1,t_1), \cdots, in(b_1,t_m),nin(b_1,t_{m+1}), \cdots,
nin(b_1,t_n)$
	\hspace*{2mm}$\elmrarrow_b$ \hspace*{2mm} $nin(b_2,t_{n+1})$

\subsection{Refinements}

It may be possible at a given time for two or more rules to become active
at the same time. The order in which the rules are handled is
nondeterministic and the handling is an atomic operation. As such, handling a
rule may lead to data consumption which could render inactive some of the
other previously active rules. By default, the $in$ and $nin$ have a
destructive behavior: once the rule activated, $in$ consumes information
and $nin$ produces information. Since the interest of the backward reading is to
provide logical links between blackboards it may not be desired to consume
existing information or produce new one. To this aim we introduce a semantic
notation in the form of \emph{square brackets} surrounding the $in$ and $nin$
primitives in order to inhibit their \emph{destructive behavior}. As such,
$[in]$ will not destroy information and $[nin]$ will not produce new information
once the rule is activated. More details will be provided in the following
section.

\subsection{Examples}
\label{sectImplLimSubExamples}
By exploiting the expressiveness of the two operational readings, we can define
special operations that would model the way in which blackboards are connected to
each-other and the way in which they react to given contexts. Let us now present
some practical uses of the rules in the context of distributed coordination.

The \emph{forward rule}, defined as
\[ \bbrdef{ forward(b_1,b_2)}
		  { \{ in(b_1,X) \elmrarrow_f in(b_2,X) \}, }
\]
would redirect the tuples destined for $b_1$ to $b_2$. This is particularly
interesting in scenarios of automatic information sharing between devices
that enter each other's communication range. Other uses may include load
balancing or for memory management. We can imagine for example that if a
blackboard has a strict memory quota limit it could use the forward rule to
bypass tuples to a remote blackboard in order to prevent overflow. A variant
can be represented as a \emph{copy rule} by simply adding brackets in the LHS:
\[ \bbrdef{ copy(b_1,b_2)}
	{ \{ [in(b_1,X)] \elmrarrow_f in(b_2,X) \}. }
\]
Every new tuple arriving on $b_1$ will be transmitted to $b_2$ without removal
from $b_1$. Used in TCP/IP, the broadcast or multicast functionalities can also
be modeled with the help of the following rule:
\[ \bbrdef{ broadcast(b,b_1,\cdots,b_n)}
	{ \{ in(b,X) \elmrarrow_f in(b_1,X),\cdots,in(b_n,X) \}. }
\]

The converse of a broadcast rule would be the merger one, with the purpose
of collecting all the tuples from several blackboards to a single one, with
particular interest in the scenarios involving sensor networks:
\[ \bbrdef{
merge(b_1,\cdots,b_n,b)} { \{ in(b_1,X)\elmrarrow_f in(b,X),\cdots,in(b_n,X) \elmrarrow_f in(b,X)\}. }
\]

The \emph{inherit rule}, defined as
\[ \bbrdef{ inherit(b_1,b_2)}
	{ \{ [in(b_1,X)] \elmrarrow_b [in(b_2,X)] \},}
\]
states in a direct form that $b_1$ inherits $b_2$ or $b_1$ has access to all
the tuples present on $b_2$. This type of rule enables a logical link
between two blackboards and defines the connection between two mobile devices.
By adding inheritance to the forward rule we allow the blackboard that is
forwarding the tuples to also keep a pointer in case it ever needs to access
those tuples again. As such, the enriched forward rule is:
\[ \bbrdef{ forward(b_1,b_2)}
	{ \{ in(b_1,X) \elmrarrow_f in(b_2,X), [in(b_1,X)] \elmrarrow_b [in(b_2,X)]\}}
\]
It should be noted that the \emph{inherit rule} must have the primitives
in the LHS and RHS guarded by square brackets in order to keep the significance
of a logical link and not to change the contents of the blackboards.

\section{Implementation}
\label{sectTechDet}

\subsection{General principles}

Being a language designed for distributed systems, \Bach\ has been
implemented in a client-server fashion, but only in terms of the
architecture. Each device acts independently and is responsible for establishing
connections with neighboring devices. There are no devices designated as
central managers for transactions. No hypothesis is made on a-priori knowledge
of the network architecture. The server-side and the client-side represent just
a separation of concepts and tasks that are to be performed on each device.

The server-side component is responsible for handling the communication, the
blackboard and its operations, a rule space for containing the reaction rules
associated with the blackboard, a request space for the agents that need to be
processed and a solved request space for storing the results of the agents
execution.

The client-side handles the parsing of string representations of agents,
dispatches requests to a local or remote blackboard depending on needs and
receives the replies. Parsing an agent returns a tree-like structure with the
nodes consisting of the composition operators that link the primitives, which
are stored in the leaves. The processing of the tree begins at the root node
with the recursive creation of sub-agents until the leaves are reached, moment
at which requests are formed towards the server.

Our purpose being to provide a fully functional framework for mobile
devices, that would support all the features of the proposed
coordination language, we found Python a good choice of programming language,
since it ensures easy portability of the code on different mobile platforms:
Symbian, iOS, Android.

\subsection{Implementation techniques}
The blackboard rules are implemented as a structure comprised of two ordered
arrays, one for the LHS and one for the RHS, and a boolean variable depicting
whether the rule is a forward reading, when true, or a backward reading, when
false.

For defining the activation condition of a rule it is not mandatory to have
a context composed solely of different tuples. Depending on the needs, several
instances of the same tuple may be required. This would translate by placing a
sequence of $in(b,t)$ primitives in the condition. In order to avoid writing
repeatedly one $in$ primitive for each needed instance of the tuple $t$ we
provide a small extension allowing to add an index to specify the total number
of instances. Subsequently, $in_c(b,t)$ states that a total count of $c$
instances of tuple $t$ are required on blackboard $b$. Testing the absence of a
tuple amounts to having 0 instances of the blackboard. Hence, no counters are
needed for the $nin$ primitives. By complementing the above definitions with
this notation we obtain the following general rules:\\
\indent $in_{c_1}(b_1,t_1), \cdots, in_{c_m}(b_1,t_m),nin(b_1,t_{m+1}),\cdots,nin(b_1,t_n)$\\
\hspace*{10mm}$\elmrarrow_f$ \hspace*{2mm} $in_{c_{n+1}}(b_{n+1},t_{n+1}),
\cdots, in_{c_p}(b_p,t_p), nin(b_{p+1},t_{p+1}), \cdots, nin(b_q,t_q)$

\indent $in_{c_1}(b_1,t_1), \cdots, in_{c_m}(b_1,t_m),nin(b_1,t_{m+1}),
\cdots, nin(b_1,t_n)$\\
\hspace*{10mm}$\elmrarrow_b$ \hspace*{2mm} $in(b_2,t_{n+1})$

\indent $in_{c_1}(b_1,t_1), \cdots, in_{c_m}(b_1,t_m),nin(b_1,t_{m+1}),
\cdots, nin(b_1,t_n)$\\
\hspace*{10mm}$\elmrarrow_b$ \hspace*{2mm} $nin(b_2,t_{n+1})$

A rule's activation condition is defined by the minimum number of tuple
instances specified in the $in$ primitives and by the absence of
tuples specified in the $nin$ primitives. Concretely, rules are handled through
an \emph{activation vector} which has the generic form:
\[
	activation~vector=[c_1,\cdots\,c_m,0,\cdots,0]
\]

At the same time, each rule must keep track of the changes occurring on the
blackboard. With the execution of $tell$ or $get$ primitives, tuples are
respectively added or removed. By keeping track of the tuples that transit the
blackboard it is possible to determine the moment when the context is met for
the rule to become active. To this purpose, a separate vector must be used in
order to count the number of tuple instances present on the blackboard
and which match the tuples defined in a rule's LHS. More specifically,
\emph{blackboard vectors} are of the form:
\[
	blackboard~vector=[bc_{t_1},\cdots,bc_{t_m},bc_{t_{m+1}},\cdots,bc_{t_n}]
\]
where $bc_{t_i}~(1 \leq i \leq m)$ represents the counter for the number of
tuples present on the blackboard that match the tuples $t_i$ of the $in$
primitives and $bc_{t_j}~(m+1 \leq j \leq n)$ represents the counter for the
number of tuples present on the blackboard that match the tuples $t_j$ of the
$nin$ primitives. This incremental style of computing allows to observe only
the modifications that occur on the blackboard. So instead of comparing the
entire LHS of a rule with the entire contents of the blackboard, only the tuple
that was used in the $tell$ or $get$ is the subject of this comparison.

By using these two definitions it is possible to formally define a rule's
activation condition:
\begin{center}
	$bc_{t_k}~\geq~c_k,~k=1\ldots~m~~~ and ~~~bc_{t_k}~=~0,~k=m+1\ldots~n$
\end{center}

It is worth to notice that in the case of the backward reading the rule stays
active as long as the above condition is met, while in the case of the forward
reading the rule acts as a reaction rule by producing or deleting tuples as
depicted by the RHS. Taking into account the definitions presented at the
beginning of this section we observe that if the activation context may be
obtained in several ways from the tuples of the blackboard, then the forward
reading rule can be executed several times. In fact, this number is
determined by the total number of combinations in which the context could be
obtained from the blackboard's content:
\[
	\prod_{k=1}^{m} {{bc_{t_k}} \choose {c_k}}
\]

This is of course a maximum number, depending on the usage of brackets on the
$in$ and $nin$ primitives, because if tuples are consumed or produced on the
blackboard due to the rule execution this may deactivate the condition of the
LHS.

\subsection{Examples}
\begin{figure}[!htbp]
	\subfigure[The evolution of the counter vector]{
		\includegraphics[width=0.4\textwidth]{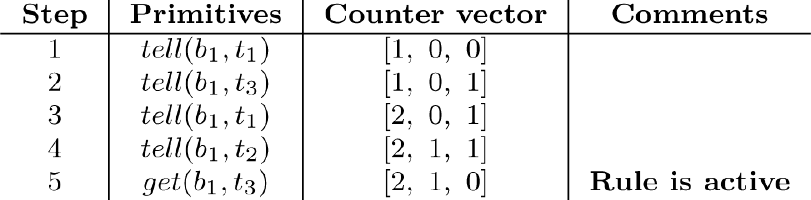}
		\label{fig:forwardReadingSub1}
	}
	\hspace{0.05\textwidth}
	\subfigure[The blackboards before and after executing the rule]{
		\includegraphics[width=0.5\textwidth]{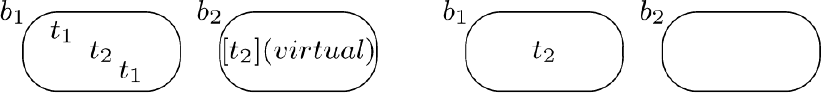}
		\label{fig:forwardReadingSub2}
	}
	\vspace{-3mm}
	\caption{Forward reading example}
	\label{fig:forwardReading}
\end{figure}

\textbf{Forward reading example}: Let us consider two blackboards, $b_1$ and
$b_2$, initially empty of tuples and the forward rule on blackboard $b_1$:
\[
	in_2(b_1,t_1),~[in(b_1,t_2)],~nin(b_1,t_3) \elmrarrow_f in(b_2,t_2)
\]
which states that if at least two instances of tuple $t_1$, one instance of
tuple $t_2$ and no instance of $t_3$ are present on blackboard $b_1$, then one
instance of tuple $t_2$ will be created on blackboard $b_2$. For this rule
the activation vector is $[2,1,0]$. Given that $b_1$ is empty, the blackboard
counter for this rule is initially $[0,0,0]$. Assuming a set of primitives
executed in the sequence depicted in figure~\ref{fig:forwardReadingSub1}, we
obtain after step 5 a state of the blackboard vector that enables the rule.

By applying the combinatorial formula, we obtain a number of three possibilities
in which the rule can be fired. However, after the first execution, two
instances of the tuple $t_1$ will be consumed from $b_1$ and one instance of the
tuple $t_3$ will be produced on $b_1$. In the meantime, due to the RHS of the
rule, one tuple $t_2$ is added to $b_2$. After this execution the state of the
two blackboards is as presented in figure~\ref{fig:forwardReadingSub2} with the
blackboard vector having the values $[1,1,1]$, which implies the deactivation
of the rule.

\begin{figure}[!htbp]
	\setlength{\unitlength}{2mm}
	\subfigure[The evolution of the counter vector]{
		\includegraphics[width=0.4\textwidth]{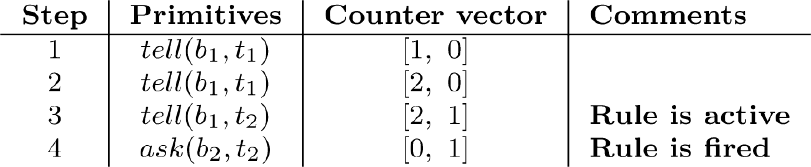}
		\label{fig:backwardReadingSub1}
	}    
	\hspace{0.05\textwidth}
	\subfigure[The blackboards before and after firing the rule]{
		\includegraphics[width=0.5\textwidth]{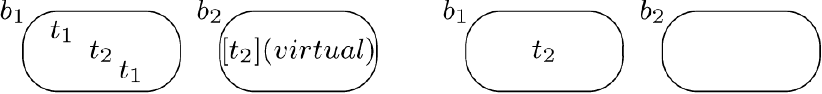}
		\label{fig:backwardReadingSub2}		
	}
	\vspace{-3mm}
	\caption{Backward reading example}
	\label{fig:backwardReading}
\end{figure}

\textbf{Backward reading example}: As with the forward reading, we use the same
two initially empty blackboards, $b_1$ and $b_2$ and associate a backward
reading rule on blackboard $b_2$:
\[
	in_2(b_1,t_1),~[in(b_1,t_2)] \elmrarrow_b [in(b_2,t_2)]
\]
having its activation vector $[2,1]$. Suppose an agent wants to execute
the primitive $ask(b_2,t_2)$. Given the fact that $b_2$ is empty, the agent
suspends its execution waiting for an instance of $t_2 $ to become
available.
In the meantime let us suppose that we execute a set of primitives on blackboard
$b_1$, as shown in figure~\ref{fig:backwardReadingSub1}. Since at step 3
the rule becomes active, blackboard $b_2$ can assume the presence of an
instance of tuple $t_2$, due to which the $ask$ primitive previously
suspended continues its execution.

By analyzing these two examples in more details we observe that in the case of
forward reading the rule needs to be associated only with the blackboard defined
in the LHS, whereas the backward reading rules require to be present on the
blackboard from the LHS as well as the RHS.

\subsection{Performance analysis}
In order to have an image on how our approach is placed with respect to other
implementations, we ran a comparative performance test on one representative of
reactive rule based model, \TUCSON.
More precisely, we have modeled the synchronization of multiple processes,
which has been presented as a practical application in the field of workflow
management~\cite{ORV06}. Assuming each process places a tuple on the blackboard to
signal that it had finished processing, the synchronization is achieved by
reacting to the presence of all the tuples on the blackboard. In our approach,
this amounts to a forward reading rule of the type:
\[
	in( \langle task,1 \rangle),~\ldots,~in( \langle task,n \rangle) \elmrarrow_f
	in( \langle task,final \rangle)
\]

In \TUCSON, since the reaction is observed only on one primitive execution, the
test scenario implies the use of $n$ rules, one for each different tuple:
\begin{center}
$reaction(out(task(1)),~completion,~(in(task(1),\ldots,in(task(n),out(task(final))))).$\\
$\ldots$\\ 
$reaction(out(task(n)),~completion,~(in(task(1),\ldots,in(task(n),out(task(final))))).$\\

\end{center}

We ran several tests, increasing the value of $n$ from 1 to 200. Two execution
steps can be identified: {\it (i)} add and parse the reaction rules, {\it (ii)}
add in sequence the tuples corresponding to the tasks: $ \langle task,1
\rangle,~\ldots,~ \langle task,n \rangle$.

\begin{figure}
\centering
	\includegraphics[width=0.45\textwidth]{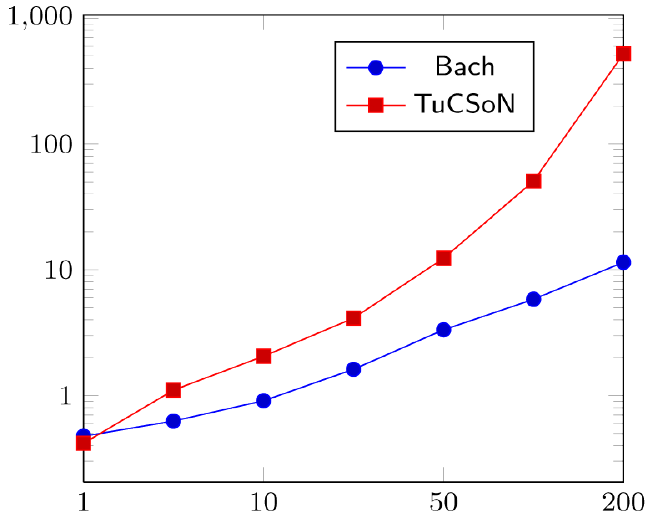}
\vspace{-5mm}
\caption{Comparative performance result}
\label{figure:results-test-1}
\end{figure}

The tests have been conducted on two different machines, one on which the tuples
1 to n were outputted and the other on which the final task was outputted. For
metrics, we have considered the total execution time. The results are plotted in
figure~\ref{figure:results-test-1} on a semilogarithmic scale where the
abscissa holds the different values of {\it n} and the ordinate the execution
time expressed in seconds.

We observe that, in most cases, up to $n=20$, the results are comparable.
The execution times differ greatly as the value of $n$ increases due to the fact
that in \TUCSON, to achieve the same functionality, there is the need to encode
$n$ rules and apply all of them each time a new tuple is added to the
blackboard.
Moreover, \TUCSON\ requires additional processing for the
transactional mechanism in the body of the reaction. On the other hand, our
approach makes use of an incremental computation in order to avoid testing the
entire rule activation condition after each operation performed on the
blackboard.

The reciprocal translation, from \TUCSON\ to \Bach , based on an example
in~\cite{ORV06} and presented in table~\ref{table:compLang}, amounts for the particular
case of $n=2$ in our test scenario.

\begin{table}[htbp]
	\centering
	\begin{tabular}{| l | l |}
		\hline
		\textbf{\TUCSON} &  \textbf{\Bach} \\
		\hline
			\lstset{columns=fullflexible,basicstyle=\ttfamily,language=Prolog}
	\begin{scriptsize}
\begin{lstlisting}
reaction(out(task_result(taskA,X),(
	in_r(task_result(taskA,X)),
	in_r(task_result(taskB,Y)),
	out_r(task_todo(taskC,args(X,Y))))).
reaction(out(task_result(taskB,Y),(
	in_r(task_result(taskB,Y)),
	in_r(task_result(taskA,X)),
	out_r(task_todo(taskC,args(X,Y))))).
\end{lstlisting}
	\end{scriptsize} 
	&
	\begin{scriptsize}
		$in(\langle taskA, ?X \rangle), in(\langle taskB, ?Y \rangle)
			\elmrarrow_f
		in(\langle taskC, !X, !Y \rangle)$
	\end{scriptsize}
	\\
	\hline

	\end{tabular}
	\caption{Syntax translation from \TUCSON\ to \Bach}
	\label{table:compLang}
	
\end{table}

Likewise, comparable results have been obtained for the examples presented in
section~\ref{sectImplLimSubExamples}. This was expected since they are
particular cases of the scenario presented above.

Based on these results, we can conclude that our approach presents
advantages in terms of performance and expressivity, mostly in scenarios which
require the verification of complex contexts.

\section{Mobile devices}
\label{sectMobCoord}
By using the rule mechanism we can model in \Bach\ the connection of remote
blackboards when a physical connection between their hosts exists. Being
given two devices A and B, the blackboard $b_A$ of host A is linked to the
blackboard $b_B$ of host B by providing the backward reading rule
$[in(b_B,X)]\longrightarrow_{b}[in(b_A,X)]$. Symmetrically, on $b_B$,
$[in(b_A,X)]\longrightarrow_{b}[in(b_B,X)]$ defines the link between the
blackboards $b_B$ and $b_A$. The disengagement is modeled by simply deleting the
rules that define the link between the blackboards.

To intermediate between the physical connection of the hosts and the logical
linking of blackboards we introduce the notion of \emph{events} that are
triggered by dedicated processes that monitor the network activity. The pair of
tuples $(connect,h)$ and $(disconnect,h)$ can be generated when the local host
connects to a remote host $h$, respectively disconnects from $h$.

Each device is responsible for the discovery of its neighbouring devices,
achieving this by use of the dedicated functionalities of the network layer. In
the same way in which Bluetooth publishes its available services to inquiring
devices, one host can publish the list of its public blackboards to other
inquiring hosts. As a result, the hosts become self-aware of the neighbouring
blackboards which could be uniquely identified by a construct of type
$blackboard\_\ name @ host\_\ name$, where the host's name can
be represented by it's MAC or IP address. We achieve the self-awareness and
self-adaptation because there is no central manager that handles the engagement
and disengagements of hosts. However, for external programming of the
blackboards an \emph{a priori} knowledge of their names is necessary. Generic
rules such as \emph{inherit}, \emph{broadcast}, \emph{merge} (introduced in
Section~\ref{sectImplLimSubExamples}) are not affected by such a restriction
since they can be kept up to date by the events mentioned in the previous
paragraph.

In mobile ad-hoc networks, sudden disconnections may occur without notice. When
such an event occurs in the course of an active rule, executing primitives on
remote blackboards which are no longer connected would fail. To prevent such an
event from blocking the local blackboard it is fair to assume a reasonable
connection timeout. This is easily achievable by using the timed primitives
introduced in~\cite{JL07a} and refined in~\cite{L07}.

As means of communication between hosts we are currently using the Bluetooth
technology in {\sf RFCOMM} mode. It is fair to say that Bluetooth
has become a standard on current smartphones. Also, if not present by default on
portable computers the low price of Bluetooth dongles makes it a viable choice
for short range communication. One drawback of Bluetooth is that devices
cannot announce their presence, but try to discover devices around them. For
this reason, the concept of two devices being \emph{``close enough''} to
establish a connection can only be modeled by periodically scanning for devices
within range.

The philosophy we aim to achieve is the one expressed by the BUMP
application for mobile devices~\cite{LHM12}. Two devices having the BUMP
application installed can be physically bumped-into each other in order to get
them connected. Once connected, it is possible to share contacts, photos, etc.
by means of drag-and-drop from one device to the other. Our experiments show
that the application works only in the presence of a functional internet
connection. Because of this requirement we concluded that BUMP could not be
classified as an application destined for mobile ad hoc networks, even if the
two devices are physically close. In response, the \Bach\ language would allow
the development of applications that would offer: support more than two devices
at once, automatic connection to compatible devices in range, local
communication without the need of an active internet connection.

From the point of view of coordination languages the mobility issue can be seen
from two perspectives: \emph{logical} and \emph{physical}, as classified in
several works~\cite{MPR01,MPR06,S02,ZME02}. The logical mobility aims at
providing to an agent (a program fragment) the means for navigating or exploring
an existing connection topology in order to pursue its execution. The physical
mobility concerns actual devices or hosts that form the nodes of a network
through which the agents roam. The latter is particularly interesting in the
case of {\sf MANETs} (Mobile ad hoc Networks) since the devices moving in and
out of each-other's range introduce a high degree of unpredictability in the
network topology. The challenge arises from the need to correctly adjust the
execution of the agents in the new hypothesis consisting of expanded or
contracted tuple spaces.

The \KLAIM\ language~\cite{NFP97} has been among the first proposals by
considering processes in the same way as data: transferable between computing
environments. Departing from the method of invoking remote procedures, \KLAIM\
allows for code fragments to be sent and executed on remote nodes by interacting
with other processes via the local tuple space. Since the transient sharing of
tuple spaces is not supported, processes need to move to new locations in order
to access new tuple spaces.

As stated in~\cite{MPR06}, in order to keep a high degree of generality at the
model level, \LIME\ ~\cite{MPR06,PMR99} does not use explicit mechanisms for
handling the mobility issue, be it logical or physical. Instead, mobility is
inferred from the changes that occur on the host-level or federated tuple spaces
as a result of an agent engaging or disengaging a group. We believe that
certain limitations are introduced by the need of having a group leader that
handles the operations of engagement and disengagement since MANETs are
characterized by highly unpredictable connections and disconnections.

\TUCSON ~\cite{OZ99}, designed as a coordination model for internet agents,
uses the notion of programmable tuple spaces and uses the hypothesis of
permanent communication links. Mobile agents roam the nodes of hierarchically
organized networks to query for information and retrieve it if found.

Because of the challenges posed by MANETs we consider that physical mobility has
a crucial impact over the logical one. In order to achieve greater flexibility
for the agent's execution, these must not be encumbered with the heterogeneity
of the surrounding blackboards possibly containing relevant information. To
support this idea we propose an alternative solution in the sense that the
blackboard on which the agent is destined to be executed should provide the
means in order to ensure connections to neighboring blackboards, which would
allow the agent faster and more transparent access to information. For this
purpose, we present in the following section a rule based mechanism that would
enable mobile agent coordination independent of the topology of the network. We
believe this to be a different point of view from the approaches presented above
in the sense of shifting the focus from handling the mobility of agents to
managing the mobility of hosts.

In this light we consider our approach as having a clear advantage with respect
to other related pieces of work. The backward reading rule in particular offers
all the necessary expressiveness in order to link remote blackboards and to provide
simple and efficient access between them.

\section{Related work}
\label{sectRelatedWork}

Let us now offer a bird's eye-view over existing lines of research related to
the idea of reactivity and see how they are placed with respect to our proposal.

\subsection{Chemical models}
The road of reactivity was paved by the \GAMMA\
model~\cite{BCM88} which introduced the idea of transforming multi-sets of data
by means of mechanisms inspired by chemistry. Accordingly, the multi-set is
metaphorically seen as a chemical solution on top of which different reaction
rules are defined. The rules consist of pairs \emph{(R,A)}, where \emph{R}
represents the reaction condition and \emph{A} the action to be taken. The
multi-set evolves as long as the reaction condition is met, after which a
stable point is reached. Data is thus rewritten producing either an expansion
or a reduction of the initial ones.

As an extension of \GAMMA , the \emph{chemical abstract machine}~\cite{BG90} or
simply \emph{cham}, added the notions of \emph{membranes} and \emph{airlocks}.
Membranes act as containers for sub-solutions, thus enforcing local reactions.
Airlocks enable the communication between these enclosed sub-solutions and their
containing environment. Such a model is of particular interest to the
coordination in mobile ad hoc networks since an analogy can be found between a
molecule and a device, respectively between an airlock and the communication
links connecting the devices.

More recent developments of the chemical metaphor are those related to
the biochemical tuple spaces~\cite{VC09}, which introduce a probabilistic
approach. By using tuple concentration and chemical rates it is possible to
model the likelihood of a given reaction occurring, service equilibrium, service
decay, service competition or service composition~\cite{VC10}. In terms of
pervasive ecosystems, the work in~\cite{VPMS12} explores how network nodes can
be enriched with {\sf live semantic annotations} which can be governed by global
{\sf eco-laws} in a chemical-like fashion. This proposition is similar to our
current one from the point of view of declarative transformations. However,
these eco-laws rely on an underlying framework covering the global space of
neighbour devices. In our approach, rules are stored on the blackboards
themselves, and connect them without other middleware. Moreover, LSA rules
always consume their reactants (lhs) and produce their products (rhs), while
ours allow a broader set of behaviours definitions.

Compared to these pieces of work, our proposal keeps the same idea of
reactivity. However, it refines it by identifying forward and backward reading,
by enhancing the patterns of the rules in distinguishing the presence or absence
of information on both sides, by providing an efficient implementation and
applying it to mobile ad hoc networks. We provide no counterpart for
probabilistic reasonings, but consider this as orthogonal to our work. As a
result, ideas from~\cite{VC09,VC10} can be introduced directly in our work. Such
ideas will be the subject of future work. 

\subsection{Reactive models}
In another line of research, the
articles~\cite{BJ96,JB01} explore models relying on the idea of reactive tuple
spaces. Among others, they are used for the coordination of mobile agents. This
has also been treated in a series of work, such as: {\sf MARS}~\cite{CLZ99},
\TUCSON\ ~\cite{OZ99}, {\sf ReSpecT} ~\cite{DNO98,DO99}, \LIME\
~\cite{MPR06,PMR99}.

More concretely, the {\sf MARS} model proposes reactions in the form of a
four components set consisting of: \emph{(i)} the reaction type, \emph{(ii)} the
tuple wild-card to be matched, \emph{(iii)} the type of operation on the tuple
space that should be monitored, \emph{(iv)} the agent's identity. This mechanism
is very flexible and is able to express a wide range of scenarios, from
specific to more general ones. This is achieved in the way in which the four
components are defined: the most general situation occurs when only the reaction
type is specified and is rendered more precise by adding values to the other
components.

In \TUCSON\ , the approach is to define programmable logic tuple centers which
consist of a tuple space enhanced by the notion of \emph{behavior
specification}. The supervision and control of this behavior is achieved by
specifying reactions to the communication events over the tuple space.
Reactions act like a set of operations handling sets of tuples in the tuple
center, either in the form of addition or deletion. Operations of
a reaction are executed atomically, in the sense that even if multiple
primitives are invoked they are perceived by the system as a single event.

\RESPECT\ implements the reactions in the form of two special types of tuples.
The first represents an association between a communication primitive and a
logical event, allowing for groups of primitives to be connected to one
identical logical event or for one primitive to generate several events. The
second is an association between the logical events and the actual reaction
body, which consists of either state primitives, term predicates or primitives
of the tuple space. The paper~\cite{O06} augments the language with the
introduction of a \emph{guard}, that may enforce additional requirements for an
event, such as its source, destination or trigger time. The multiple blackboard
approach for \RESPECT\ has been explored in~\cite{MOV02} by studying its
interactions with \LOGOP, introduced in~\cite{SM02} and later refined
in~\cite{MOV04}. \LOGOP\ presents itself as an extension of \LINDA\ for the
management of multiple tuple space environments. The execution of primitives in
such a hypothesis is supported by the introduction of composed tuple spaces in
their definition. The composition is achieved by applying the logical operators
{\sc AND}, {\sc OR}, {\sc XOR}, {\sc NOT} between multiple tuple spaces. A
dedicated {\sf logOp} tuple center uses the \RESPECT\ language to react to
\LOGOP\ primitives and form requests for each tuple center forming the composition. It
also receives the replies and forms the final answer. We believe that this
architecture, using a centralised manager, is not suitable in the unpredictable
scenario of mobile ad-hoc networks.

Offering a different perspective from the previous models, \LIME\
associates reactions to the context of the tuple space rather than
the set of primitives executed on it. Reactions are triggered by matching
tuples on the tuple space with given patterns, thus defining specific contexts.

To sum up, the main characteristics of related work are threefold: \emph{(i)}
the reaction condition is expressed only in terms of data being present on the
blackboard, \emph{(ii)} some reactions are triggered on the execution
of primitives, \emph{(iii)} the reaction rules mostly concern a single
blackboard.

In our approach we provide a finer control over the reaction conditions, which
can be defined in terms of data presence (by using the primitive called
\emph{in}) and data absence (by using the primitive called \emph{nin}). This
allows the specification of more precise and strict contexts, not possible in
other works. In addition, our rule mechanism is designed to be used also in the
multi-blackboard system proper to mobile ad hoc networks.

\subsection{Final remarks}
In summary, as may be appreciated by the reader, our work presents significant
differences with respect to the related work. In addition to the chemical
models, our approach allows the definition of more complex contexts consisting
not only of information that needs to be present, but also of information that
needs to be absent. In the terms of the chemical metaphor, our proposition
offers the possibility to model the idea of an \emph{inhibitor}, a substance
capable of stopping or retarding a chemical reaction. By means of the $nin$
primitive it is possible to express reactions which occur in the absence of
given inhibitors.

With respect to the reactive models, in which the main focus is put on reacting
to the execution of atomic primitives, our declarative approach, by means of
context-awareness, offers greater flexibility and expressiveness in terms of
coordinating mobile agents in mobile environments and in offering the necessary
means to ensure logical links between remote blackboards. In particular, our
blackboard rules allow for a fine grained tuning of actions to be taken when
mobile devices come close enough. In addition to the merge of tuple spaces,
offered by languages such as \LIME, we are able to code alternative forms of
coordination like \emph{filtering} or \emph{duplication} of blackboard contents.

\section{Conclusions}
\label{sectConcl}

In this paper, we propose a solution for coordination in mobile ad hoc
networks by means of declarative rules. By providing two operational readings,
we aim at rendering a double functionality. On the one hand, by means of
\emph{forward reading}, we offer support for context awareness and reactivity
to context changes. By modelling the user's needs as contexts, it is possible to
react promptly once the conditions are met. On the other hand, by means of
\emph{backward reading}, we provide a mechanism for supporting the transient
sharing of blackboards of mobile devices that come within communication range.
The \emph{backward reading} rules build logical links between remote
blackboards, which enables their unification without the need to physically
replicate the tuples. In conjunction, the \emph{forward reading} and
\emph{backward reading} support the possibility of interacting not only with the
local blackboard, but with remote ones as well.

Certainly, the idea of rules and reactivity is not a novelty to coordination
models. However, we have introduced new variants based on a declarative reading
which we have shown to be expressive, yet being efficiently implementable.
Taking an incremental approach on the computation bypasses the need for a
transactional mechanism, since keeping track of how the blackboard content is
reflected on the rule's activation condition there is no need to reevaluate it
after each primitive execution.

For future work we aim at introducing a probabilistic reasoning in the sense of
attaching reaction rates to the rules, similar to~\cite{VC09}. By exploiting
the rules mechanism we also aim to obtain capabilities for complex event
processing.

\bibliographystyle{eptcs}
\bibliography{references}

\end{document}